\documentclass[preprint,showpacs,preprintnumbers,amsmath,amssymb]{revtex4}

\newcommand{\ds}{\displaystyle}

\usepackage{graphicx}
\usepackage{dcolumn}
\usepackage{bm}

\newcommand{\be}{\begin{equation}}
\newcommand{\en}{\end{equation}}
\newcommand{\bea}{\begin{eqnarray}}
\newcommand{\ena}{\end{eqnarray}}

\begin{document}


\title{Emergent Universe by Tunneling}

\author{Pedro Labra\~{n}a}
 \email{plabrana@ubiobio.cl}
\affiliation{Departamento de F\'{\i}sica, Universidad del
B\'{\i}o-B\'{\i}o, Avenida Collao 1202, Casilla 5-C, Concepci\'on,
Chile.}

\date{\today}

\begin{abstract}

In this work we propose an alternative scheme for an Emergent
Universe scenario where the universe is initially in a static state
supported by a scalar field located in a false vacuum. The universe
begins to evolve when, by quantum tunneling, the scalar field decays
into a state of true vacuum.
The Emergent Universe models are interesting since they provide
specific examples of non–singular inflationary universes.

\end{abstract}

\maketitle

\section{Introduction}
\label{Int}

Cosmological inflation has become an integral part of the standard
model of the universe. Apart from being capable of removing the
shortcomings of the standard cosmology, it gives important clues for
large scale structure formation. The scheme of inflation \cite{Guth,
Albrecht, Linde1, Linde2} (see \cite{libro} for a review) is based
on the idea that there was a early phase in which the universe
evolved through accelerated expansion in a short period of time at
high energy scales. During this phase, the universe was dominated by
the potential $V(\phi)$ of a scalar field $\phi$, which is called
the inflaton.

 Singularity theorems have been devised that apply in the
inflationary context, showing that the universe necessarily had a
beginning (according to classical and semi-classical theory)
\cite{Borde:1993xh,Borde:1997pp, Borde:2001nh,
Guth:1999rh,Vilenkin:2002ev}. In other words, according to these
theorems, the quantum gravity era cannot be avoided in the past even
if inflation takes place.
However, recently, models that escape this conclusion has been
studied in Refs.
\cite{Ellis:2002we,Ellis:2003qz,Mulryne:2005ef,Mukherjee:2005zt,
Mukherjee:2006ds,Banerjee:2007qi,Nunes:2005ra,Lidsey:2006md}. These
models do not satisfy the geometrical assumptions of these theorems.
Specifically, the theorems assume that either {\bf i)} the universe
has open space sections, implying $k = 0$ or $-1$, or \textbf{ii)}
the Hubble expansion rate $H$ is bounded away from zero in the past,
$H > 0$.

In particular, Refs.
\cite{Ellis:2002we,Ellis:2003qz,Mulryne:2005ef,Mukherjee:2005zt,
Mukherjee:2006ds,Banerjee:2007qi,Nunes:2005ra,Lidsey:2006md}
consider closed models in which $k = +1$ and $H$ can become zero, so
that both assumptions \textbf{i)} and \textbf{ii)} of the
inflationary singularity theorems are violated.
In these models the universe is initially in a past eternal
classical Einstein static (ES) state which eventually evolves into a
subsequent inflationary phase.
Such models, called Emergent Universe, are appealing since they
provide specific examples of non–singular (geodesically complete)
inflationary universes.

Normally in the Emergent Universe scenario, the universe is
positively curved and initially it is in a past eternal classical
Einstein static state which eventually evolves into a subsequent
inflationary phase, see
\cite{Ellis:2002we,Ellis:2003qz,Mulryne:2005ef,Mukherjee:2005zt,
Mukherjee:2006ds,Banerjee:2007qi,Nunes:2005ra,Lidsey:2006md}.

\begin{figure}
\centering
\includegraphics[width=10cm]{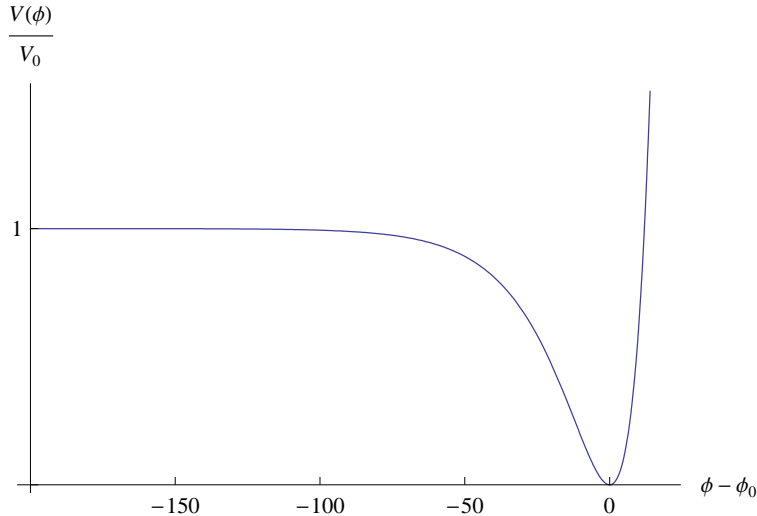}
\caption{Schematic representation of a potential for a standard
Emergent Universe scenario. \label{Potencial-1}}
\end{figure}

For example, in the original scheme \cite{Ellis:2002we,
Ellis:2003qz}, it is assumed that the universe is dominated by a
scalar field (inflaton) $\phi$ with a scalar potential $V(\phi)$
that approach a constant $V_0$ as $\phi \rightarrow -\infty$ and
 monotonically rise once  the scalar field exceeds a
certain value $\phi_0$, see Fig. (\ref{Potencial-1}).

During the past-eternal static regime it is assumed that the scalar
field is rolling on the asymptotically flat part of the scalar
potential with a constant velocity, providing the conditions for a
static universe. But once the scalar field exceeds some value, the
scalar potential slowly droops from its original value. The overall
effect of this is to distort the equilibrium behavior breaking the
static solution.
If the potential has a suitable form in this region, slow-roll
inflation will occur, thereby providing a 'graceful entrance' to
early universe inflation.

This scheme for a Emergent Universe have been used not only on
models based on General Relativity \cite{Ellis:2002we,
Ellis:2003qz}, but also on models where non-perturbative quantum
corrections of the Einstein field equations are considered
\cite{Mulryne:2005ef,Nunes:2005ra, Lidsey:2006md}, in the context of
a scalar tensor theory of gravity \cite{delCampo:2007mp,
delCampo:2009kp} and recently in the framework of the so-called two
measures field theories \cite{delCampo:2010kf,delCampo:2011mq,
Guendelman:2011fq,Guendelman:2011fr}.
%


Another possibility for the Emergent Universe scenario is to
consider models in which the scale factor asymptotically tends to a
constant in the past \cite{Mukherjee:2005zt, Mukherjee:2006ds,
Banerjee:2007sg,Debnath:2008nu, Paul:2008id, Beesham:2009zw,
Debnath:2011qi, Mukerji:2011wq}.

We can note that both schemes for a Emergent Universe are not truly
static during the static regime. For instance, in the first scheme
during the static regime the scalar field  is rolling on the flat
part of its potential. On the other hand, for the second scheme the
scale factor is only asymptotically static.

\begin{figure}
\centering
\includegraphics[width=10cm]{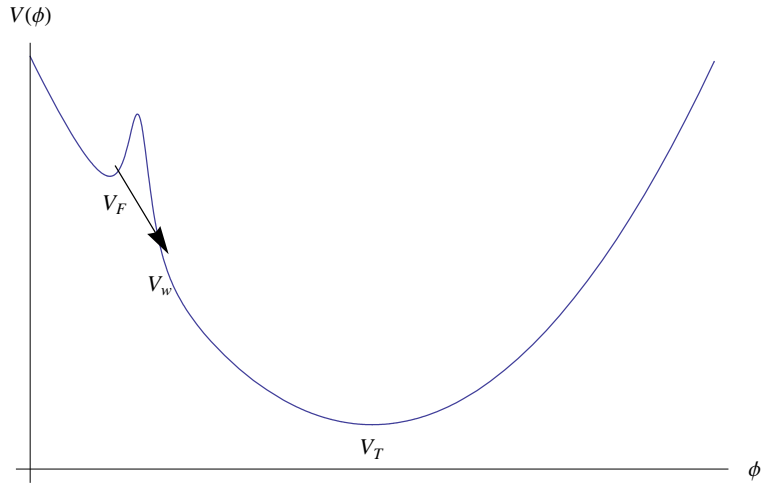}
\caption{A double-well inflationary potential $V(\phi)$. In the
graph, some relevant values  are indicated. They are the false
vacuum $V_F=V(\phi_F)$ from which the tunneling begins,
$V_W=V(\phi_W)$ where the tunneling stops and where the inflationary
era begins, while $V_T=V(\phi_T)$ denote the true vacuum energy.
\label{Potential-2}}
\end{figure}

In this paper we propose an alternative scheme for an Emergent
Universe scenario, where the universe is initially  in a truly
static state. This state is supported by a scalar field which is
located in a false vacuum ($\phi = \phi_F$), see
Fig.(\ref{Potential-2}). The universe begins to evolve when, by
quantum tunneling, the scalar field decays into a state of true
vacuum. Then, a small bubble of a new phase of field value $\phi_W$
can form, and expand as it converts volume from high to low vacuum
energy and feeds the liberated energy into the kinetic energy of the
bubble wall. This process was first studied by Coleman and Coleman
\& De Luccia in \cite{Coleman:1977py, Coleman:1980aw}.

Inside the bubble, space-like surfaces of constant $\phi$ are
homogeneous surfaces of constant negative curvature. One way of
describing this situation is to say that the interior of the bubble
always contains an open Friedmann-Robertson-Walker universe
\cite{Coleman:1980aw}.
If the potential has a suitable form, inflation and reheating may
occur in the interior of the bubble as the field rolls from $\phi_W$
to the true minimum at $\phi_T$, in a similar way to what happens in
models of Open Inflationary Universes, see for example \cite{linde,
re8, delC1, delC2, Balart:2007je}.


\begin{figure}
\centering
\includegraphics[width=10cm]{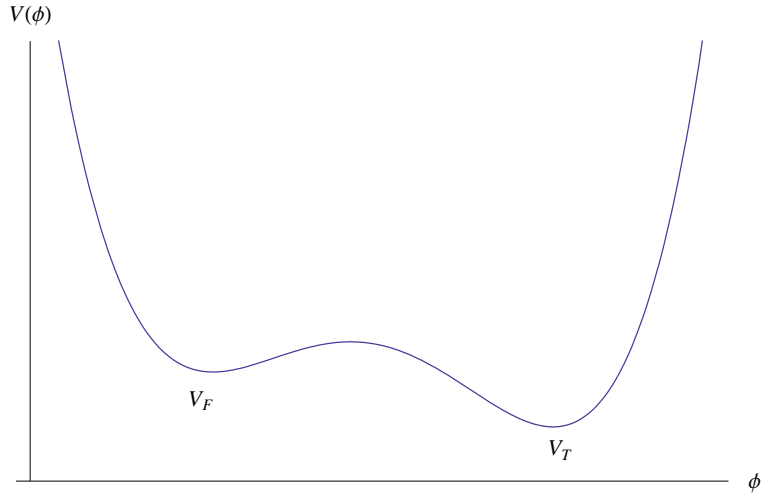}
\caption{Potential with a false and true vacuum.
\label{Potential-3}}
\end{figure}

The advantage of this scheme (and of the Emergent Universe in
general), over the Eternal Inflation scheme is that it correspond to
a realization of a singularity-free inflationary universe.
In fact, Eternal Inflation is usually future eternal but it is not
past eternal, because in general space-time that allows for
inflation to be future eternal, cannot be past null complete
\cite{Borde:1993xh,Borde:1997pp, Borde:2001nh,
Guth:1999rh,Vilenkin:2002ev}. On the other hand Emergent Universe
are geodesically complete.

Notice that in our scheme for an Emergent Universe, the metastable
state which support the initial static universe could exist only a
finite amount of time. In our scheme of Emergent Universe, the
principal point is not that the universe could have existed an
infinite period of time, but that in our model the universe is
non-singular because the background where the bubble materializes is
geodesically complete. This implies that we have to consider the
problem of the initial conditions for a static universe. Respect to
this point, there are very interesting possibilities discussed for
example in the early works on EU, see Ref.~\cite{Ellis:2003qz}.

One of these options is to explore the possibility of an Emergent
Universe scenario within a string cosmology context, where it have
been shown that the Einstein static universe is one of only two
asymptotic solutions of the Ramond-Ramond sector of superstring
cosmology \cite{Antoniadis}.

Other possibility is that the initial Einstein static universe is
created from \textit{nothing} \cite{Tryon,Vilenkin-cre}. With
respect to this, recently the possibility of a static universe
created from \textit{nothing} have been discussed by Vilenkin in
Ref.~\cite{Mithani:2011en}, where it is shown an explicit example.

The study of the Einstein Static (ES) solution as a preferred
initial state for our universe have been considered in the past,
where it has been proposed that entropy considerations favor the ES
state as the initial state for our universe, see
\cite{Gibbons:1987jt,Gibbons:1988bm}.

In this paper we consider a simplified version of the Emergent
Universe by tunneling, where the focus is on studying the process of
creation and evolution of a bubble of true vacuum in the background
of an ES universe.

This is motivated because we are mainly interested in the study of
new ways of leaving the static period and begin the inflationary
regime in the context of Emergent Universe models.

In particular, in this paper we consider an inflaton potential
similar to Fig. (\ref{Potential-3}) and study the process of
tunneling of the scalar field from the false vacuum $\phi_F$ to the
true vacuum $\phi_T$ and the consequent creation and evolution of a
bubble of true vacuum in the background of an ES universe.

The simplified model studied here contains the essential elements of
the scheme we want to present (EU by tunneling), so we postpone the
detailed study of the inflationary period, which occurs after the
tunneling, for future work.

Nevertheless, given the similarities, we expect that the behavior
inside the bubble of the non-simplified version of the EU by
tunneling will be similar to the models of single-field open
inflation. Then, if inflation inside the bubble is long, the
universe will be almost exactly flat, see \cite{Gott:1982zf,
Gott:1984ps, Sasaki:1993ha, Ratra:1994vw, Bucher:1994gb,
Sasaki:1994yt, Lyth:1995cw, Yamamoto:1995sw, Bucher:1995ga, linde,
re8, delC1, delC2, Balart:2007je}.
The density perturbation arise in these scheme in the usual way by
the quantum fluctuation of the scalar field (inflaton) as the field
slow-rolls to the true minimum. With respect to this, the general
formula for the power spectrum for the single-field open inflation
was given in \cite{Garriga:1997wz, Garriga:1998he}. Recently, in the
context of the string landscape, the contribution to the CMB
anisotropies of the perturbation in the open inflation scenario was
study in \cite{Yamauchi:2011qq}.
The detailed study of these topics is beyond the scope of this work,
but we expect to return to these points in the near future.

The paper is organized as follow. In Sect.~\ref{Sec-static} we study
a Einstein static universe supported by a scalar field located in a
false vacuum. In Sect.~\ref{Sec-Bubble}  we study the tunneling
process of the scalar field from the false vacuum to the true vacuum
and the consequent creation of a bubble of true vacuum in the
background of Einstein static universe.
In Sect.~\ref{Sec-evolution} we study the evolution of the bubble
after its materialization.
In Sect.~\ref{Sec-conclusions} we summarize our results.

\section{Static Universe Background}\label{Sec-static}

Based on the standard Emergent Universe (EU) scenario, we consider
that the universe is positively curved and it is initially in a past
eternal classical Einstein static state. The matter of the universe
is modeled by a standard perfect fluid $P=(\gamma -1)\rho$ and a
scalar field (inflaton) with energy density $\rho_\phi =
\frac{1}{2}(\partial_t\phi)^2 + V(\phi)$ and pressure $P_\phi =
\frac{1}{2}(\partial_t\phi)^2 - V(\phi)$. The scalar field potential
$V(\phi)$ is depicted in Fig.~\ref{Potential-3}. The global minimum
of $V(\phi)$ is tiny and positive, at a field value $\phi_T$, but
there is also a local false minimum  at $\phi=\phi_F$.

We have consider that the early universe is dominated by two fluids
because in our scheme of the EU scenario, during the static regime
the inflaton remains static at the false vacuum, in contrast to
standard EU models where the scalar field rolls on the
asymptotically flat part of the scalar potential. Then, in order to
obtain a static universe we need to have another type of matter
besides the scalar field.
For this reason we have included a standard perfect fluid.
For simplicity we are going to consider that there are no
interactions between the standard perfect fluid and the scalar
field.

The metric for the static state is given by the closed
Friedmann-Robertson-Walker metric:

 \be \ds
d{s}^{2}\,=\, d{t}^{2}\,-\,
a(t)^{2}\left[\frac{dr^2}{1-\frac{r^2}{R^2}} + r^2\,(d\theta^2 +
\sin^2 \!\!\theta\,\, d\phi^2 )\right], \label{met1} \en
where $a(t)$ is the scale factor, $t$ represents the cosmic time and
the constant $R>0$.
We have explicitly written $R$ in the metric in order to make more
clear the effects of the curvature on the bubble process
(probability of creation and propagation of the bubble).

Given that there are no interactions between the standard fluid and
the scalar field, they separately obey energy conservation and
Klein– Gordon equations,
\begin{eqnarray}
\partial_t \rho + 3\gamma \,H\,\rho = 0 \label{rho}\,,\\
\nonumber \\
\partial^2_t\phi+3H\,\partial_t\phi=-\frac{\partial
V(\phi)}{\partial\phi}\,,
\label{phi}
\end{eqnarray}
where $H = \partial_t a /a$.

The Friedmann and the Raychaudhuri field equations become,
\begin{eqnarray}
H^2 = \frac{8\pi G}{3}\left(\rho + \frac{1}{2}(\partial_t\phi)^2 +
V(\phi)
\right) - \frac{1}{R^2 a^2}, \label{H} \\
\nonumber\\
\partial^2_t a = - \frac{8\pi
G}{3}\,a\left[\left(\frac{3}{2}\gamma -1\right)\rho + \dot{\phi}^2 -
V(\phi) \right]\label{H2}.
\end{eqnarray}

The  static universe is characterized by the conditions
$a=a_0=Cte.$, $\partial_ta_0=\partial^2_ta_0=0$ and
$\phi=\phi_F=Cte.$, $V(\phi_F) = V_F$ corresponding to the false
vacuum.

From Eqs.~(\ref{rho}) to (\ref{H2}), the static solution for a
universe dominated by a scalar field placed in a false vacuum and a
standard perfect fluid, are obtained if the following conditions are
met
\begin{eqnarray}
\rho_0 &=& \frac{1}{4\pi G}\,\frac{1}{\gamma\,R^2 a^2_0}\,, \\
\nonumber \\
 V_F &=& \left(\frac{3}{2}\gamma -1\right)\rho_0 \,,
\end{eqnarray}
where $\rho_0$ is energy density of the perfect fluid present in the
static universe. Note that $\gamma > 2/3$ in order to have a
positive scalar potential.

By integrating Eq.~(\ref{rho}) we obtain
\begin{equation}
\rho = \frac{A}{a^{3\gamma}}\,,
\end{equation}
where $A$ is an integration constant.
By using this result, we can rewrite the conditions for a static
universe as follow
\begin{eqnarray}
A = \frac{1}{4\pi G}\;\frac{a^{3\gamma-2}_0}{\gamma\,R^2}\,,\\
\nonumber \\
V_F = \left(\frac{3}{2}\gamma -1\right)\frac{1}{4\pi
G}\,\frac{1}{\gamma\,R^2 a^2_0}\,.
\end{eqnarray}

In a purely classical field theory if the universe is static and
supported by the scalar field located at the false vacuum $V_F$,
then the universe remains static forever. Quantum mechanics makes
things more interesting because the field can tunnel through the
barrier and by this process create a small bubble where the field
value is $\phi_T$. Depending of the background where the bubble
materializes, the bubble could expanded or collapsed
\cite{Fischler:2007sz, Simon:2009nb}.

\section{Bubble Nucleation}\label{Sec-Bubble}

In this section we study the tunneling process of the scalar field
from the false vacuum to the true vacuum and the consequent creation
of a bubble of true vacuum in the background of Einstein static
universe.

Given that in our case the geometry of the background correspond to
a Einstein static universe and  not a de Sitter space, we proceed
following the scheme developed in
\cite{KeskiVakkuri:1996gn,Simon:2009nb}, instead of the usual
semiclassical calculation of the nucleation rate based on instanton
methods \cite{Coleman:1980aw}.

In particular, we will consider the nucleation of a spherical bubble
of true vacuum $V_T$   within the false vacuum $V_F$. We will assume
that the layer which separates the two phases (the wall) is of
negligible thickness compared to the size of the bubble (the usual
thin-wall approximation). The energy budget of the bubble consists
of latent heat (the difference between the energy densities of the
two phases) and surface tension.

In order to eliminate the problem of  predicting the reaction of the
geometry to an essentially a-causal quantum jump, we neglect during
this computation the gravitational back-reaction of the bubble onto
the space-time geometry.

The gravitational back-reaction of the bubble will be consider in
the next chapter when we study the evolution of the bubble after its
materialization.

In our case the shell trajectory  follows from the action (see
\cite{KeskiVakkuri:1996gn, Basu:1991ig})
\begin{equation}\label{accion}
S = \int dy \Bigg\{ 2\pi\,\epsilon\, \bar{a}_0^4\Big[\chi -
\cos(\chi)\sin(\chi)\Big] - 4\pi\,\sigma \,\bar{a}_0^3
\,\sin^2(\chi)\sqrt{1- {\chi'}^2}\Bigg\}.
\end{equation}
where we have denoted the coordinate radius of the shell as $\chi$,
and we have written the static ($a = a_0 = Cte.$) version of the
metric Eq.(\ref{met1}) as
\begin{equation}
\label{met2} ds^2 = \bar{a}_0^2\Big(dy^2 - d\chi^2 -
\sin^2(\chi)d\Omega^2 \Big),
\end{equation}
with $\frac{r}{R} = \sin(\chi)$, $\bar{a}_0 = R\,a_0$, $dt =
\bar{a}_0\,dy$ and prime means derivatives respect to $y$.

In the action  (\ref{accion}), $\epsilon$ and $\sigma$ denote,
respectively, the latent heat and the surface energy density
(surface tension) of the shell.

The action (\ref{accion}) describes the classical trajectory of the
shell after the tunneling. This trajectory emanates from a classical
turning point, where the canonical momentum
\begin{equation}
\label{PP} P = \frac{\partial S}{\partial \chi'} = 4\pi\,\sigma\,
\bar{a}_0^3\,\chi'\,\frac{\sin^2(\chi)}{\sqrt{1- {\chi'}^2}}\,,
\end{equation}
vanishes \cite{KeskiVakkuri:1996gn}.
In order to consider tunneling, we evolve this solution back to the
turning point, and then try to shrink the bubble to zero size along
a complex $y$ contour, see \cite{KeskiVakkuri:1996gn,Simon:2009nb}.
For each solution, the semiclassical tunneling rate is determined by
the imaginary part of its action, see \cite{KeskiVakkuri:1996gn}:
\begin{equation}
\label{P} \Gamma \approx e^{-2Im [S]}\,.
\end{equation}

From the action (\ref{accion}) we found the equation of motion
\begin{equation}
\label{ecmov} \frac{\sin^2(\chi)}{\sqrt{1-{\chi'}^2}} =
\frac{\epsilon \, \bar{a}_0}{2 \sigma} \Big[\chi -
\cos(\chi)\sin(\chi)\Big].
\end{equation}

The action (\ref{accion}) can be put in a useful form by using
Eq.(\ref{ecmov}), and changing variables to $\chi$:

\begin{equation}
S = \int d\chi
\,\frac{4\pi}{3}\,\epsilon\,a_0^4\,\sin^2(\chi)\sqrt{\left(\frac{3[\chi
- \cos(\chi)\sin(\chi)]}{2\sin^2(\chi)}\right)^2 -\, \bar{r}_0^2}
\;,
\end{equation}
where $\bar{r}_0 = \frac{r_0}{R}$ and $r_0 = \frac{3
\sigma}{\epsilon \,a_0}$ is the radio of nucleation of the bubble
when the space is flat ($R \rightarrow \infty)$ and static (i.e.
when the space is Minkowsky).

The nucleation radius $\bar{\chi}$ (i.e. the coordinate radius of
the bubble at the classical turning point), is a solution to the
condition $P =0$. Then from Eq.~(\ref{PP}) we obtain
\begin{equation}
\label{radio} \frac{\bar{\chi} -
\cos(\bar{\chi})\sin(\bar{\chi})}{\sin^2(\bar{\chi})} = \frac{2
\sigma}{\epsilon \, \bar{a}_0}.
\end{equation}

The action (\ref{accion}) has an imaginary part coming from the part
of the trajectory $0 < \chi < \bar{\chi}$, when the bubble is
tunneling:
\begin{equation}\label{im-s}
Im[S] = \frac{4\pi}{3}\,\epsilon\,a_0^4\,\int^{\bar{\chi}}_0 d\chi
\,\sin^2(\chi)\sqrt{\bar{r}_0^2 -\, \left(\frac{3[\chi -
\cos(\chi)\sin(\chi)]}{2\sin^2(\chi)}\right)^2} \;,
\end{equation}

Expanding (\ref{im-s}) at first nonzero contribution in $\beta=
(r_0/R)^2$ we find
\begin{equation}
\label{im-s1} Im[S] = \frac{27\,\sigma^4\,\pi}{4\,\epsilon^3}\Big[1
- \frac{1}{2}\beta^2 \Big]
\end{equation}

This result is in agreement with the expansion obtained in
\cite{Abbott:1987xq}.
Then, the nucleation rate is
\begin{equation}
\Gamma \approx e^{-2Im S} \approx
 \exp\left[-\frac{27 \sigma^4 \pi}{2
\epsilon^3} \left(1 - \frac{9\sigma^2}{2\epsilon^3\,a_0^2R^2}
\right) \right].
\end{equation}

We can note that the probability of the bubble nucleation is
enhanced by the effect of the curvature of the closed static
universe background.

\section{Evolution of the Bubble}\label{Sec-evolution}

In this section we study the evolution of the bubble after the
process of tunneling.
During this study we are going to consider the gravitational
back-reaction of the bubble.
We follow the approach used in \cite{Fischler:2007sz} where it is
assumed that the bubble wall separates space-time into two parts,
described by different metrics and containing different kinds of
matter.
The bubble wall is a timelike, spherically symmetric hypersurface
$\Sigma$, the interior of the bubble is described by a de Sitter
space-time  and the exterior by the static universe discussed in
Sec.~\ref{Sec-static}. The Israel junction conditions
\cite{Israel:1966rt} are implemented in order to join these two
manifolds along their common boundary $\Sigma$. The evolution of the
bubble wall is determined by implementing these conditions.

We will follow the scheme and notation of \cite{Fischler:2007sz}.
Then, Latin and Greek indices denote 3-dimensional objects defined
on the shell and 4-dimensional quantities, respectively. The
projectors are $e^\alpha_a = \frac{\partial x^\alpha}{\partial y^a}$
and semi-colon is shorthand for the covariant derivative.
Unit as such that $8\pi \,G = 1$.

In particular, the exterior of the bubble is described by the metric
Eq.~(\ref{met1}) and the equations (\ref{rho}-\ref{H2}), previously
discussed in Sec.~\ref{Sec-static}. At the end, the static solution
for these equations will be assumed. The interior of the bubble will
be described by the metric of the \mbox{de Sitter} space-time in its
open foliation, see \cite{Coleman:1980aw}

\begin{equation}\label{metrica-interior}
ds^2 = dT^2 - b^2(T)\left( \frac{dz^2}{1+z^2} + z^2\,d\Omega_2
\right),
\end{equation}

where the scale factor satisfies

\begin{equation}
\left(\frac{db}{dT} \right)^2 = \left(\frac{V_T}{3}\right) b^2(T) +1
\,.
\end{equation}

These two regions are separated by the bubble wall $\Sigma$, which
will be assumed to be a thin-shell and spherically symmetric.
Then, the intrinsic metric on the shell is \cite{Berezin:1987bc}
\begin{equation}\label{metrica-burbuja}
ds^2|_\Sigma = d\tau^2 - B^2(\tau)\,d\Omega_2 \,,
\end{equation}
where $\tau$ is the shell proper time.

Know we proceed to impose the Israel conditions in order to joint
the manifolds along there common boundary $\Sigma$.
The first of Israel's conditions impose that the metric induced on
the shell from the bulk 4-metrics on either side should match, and
be equal to the 3-metric on the shell.
Then by looking from the outside to the bubble-shell we can
parameterize the coordinates  $r = x(\tau)$ and $t = t(\tau)$,
obtaining the following match conditions, see \cite{Fischler:2007sz}

\begin{equation}
a(t)x = B(\tau)\,,\;\;\;\;\;
\left(\frac{dt}{d\tau} \right)^2 = 1 + \frac{a(t)^2}{1-
\left(\frac{x}{R}\right)^2}\left(\frac{dx}{d\tau}\right)^2\,,\label{cond1}
\end{equation}
where all the variables in these equations are thought as functions
of $\tau$.
On the other hand, the  angular coordinates of metrics (\ref{met1})
and (\ref{metrica-burbuja}) can be just identified in virtue of the
spherical symmetry.

The second junction condition could be written as follow
\begin{equation}\label{segunda-condicion}
[K_{ab}] - h_{a b}[K] = S_{ab},
\end{equation}
where $K_{ab}$ is the extrinsic curvature of the surface $\Sigma$
and square brackets stand for discontinuities across the shell.
Following \cite{Fischler:2007sz}, we assume that the surface
energy-momentum tensor $S_{ab}$ has a perfect fluid form given by
${S_\tau}^\tau \equiv \sigma$ and ${S_\theta}^\theta = {S_\phi}^\phi
\equiv -\bar{P}$, where $\bar{P} = (\bar{\gamma} -1)\sigma$.
Also, because of the spherical symmetry and the form of the metric
Eq.~(\ref{metrica-burbuja}), the extrinsic curvature ${K_a}^b$ has
only independent components ${K_\tau}^\tau$ and ${K_\theta}^\theta =
{K_\phi}^\phi$.
Then, from the second junction condition we obtain the following
independent equations
\begin{eqnarray}\label{condition}
-\frac{\sigma}{2} &=& [K^\theta_\theta],\\
\bar{P} &=& [K^\tau_\tau] + [K^\theta_\theta],
\end{eqnarray}
where $\sigma$ and $\bar{P}$ are considered as purely functions of
$\tau$.
Also, the junctions conditions imply a conservation law
\cite{Berezin:1987bc}, which in this case take the following form
\begin{equation}
\label{cons2} \frac{d\sigma}{d\tau} + \frac{2}{B}\frac{d
B}{d\tau}\,(\sigma + P) + [T^n_\tau]=0\,,
\end{equation}
where
\begin{equation}\label{T}
[T^n_\tau] = (e^\alpha_\tau \,T^\beta_\alpha \,n_\beta )_{out} -
(e^\alpha_\tau \,T^\beta_\alpha \,n_\beta )_{in}\,\,,
\end{equation}
and $n_\alpha$ is the outward normal vector to the surface $\Sigma$.

The evolution of the shell is completely determined by
Eq.~(\ref{condition}) and Eq.~(\ref{cons2}).
Following \cite{Fischler:2007sz} we write these matching conditions
in terms of the outside coordinates.

The extrinsic curvature could be written as:
\begin{equation}
K_{ab} = n_{\alpha ; \beta}\, e^\alpha_a \,e^\beta_b \,.
\end{equation}

The projectors of the static side are:
\begin{eqnarray}
u^\alpha \equiv e^\alpha_\tau = \left( \frac{dt}{d\tau},
\frac{dx}{d\tau}, 0, 0\right),\\
\nonumber \\
e^\alpha_\theta = (0,0,1,0), \;\;\;\; e^\alpha_\phi = (0,0,0,1).
\end{eqnarray}

We can note that $u^\alpha$ is the 4-velocity of the bubble-shell.
Then we obtain
\begin{equation}
n_\alpha = \frac{a}{\sqrt{1 -
(\frac{x}{R})^2}}\left(-\dot{x},\dot{t}, 0 ,0 \right),
\end{equation}
where dots means differentiation with respect to $\tau$ and we have
used the following conditions $u^\alpha n_\alpha = 0$ and $n^\alpha
n_\alpha = -1$, in order to determinate $n_\alpha$.

Then $K^\theta_\theta$ on the static side becomes
\begin{equation}\label{K-out}
K^\theta_{\theta (out)} = \left( \frac{a\,x\dot{x}\,a,_t \,+\, (1 -
x^2/R^2)\dot{t}}{B \sqrt{1 - (\frac{x}{R})^2}}\right).
\end{equation}

Repeating the above calculation for $K^\theta_\theta$ on the inside
we obtain
\begin{equation}\label{K-in}
K^\theta_{\theta (in)} = \left(\frac{z\,b\frac{db}{dT}\,\dot{z} +
(1+z^2)\dot{T}}{B \sqrt{1+z^2}}\right).
\end{equation}

By using Eq.~(\ref{K-out}) and Eq.~(\ref{K-in}) we can obtain the
explicit form of the junction condition Eq.~(\ref{condition}).
Nevertheless, it is most convenient write this condition as follow,
see \cite{Berezin:1987bc, Fischler:2007sz},
\begin{equation}\label{D0}
\sqrt{\dot{B}^2 - \Delta_{out}} - \sqrt{\dot{B}^2 - \Delta_{in}} =
-\frac{\sigma B}{2}\,.
\end{equation}

Where we have defined
\begin{eqnarray}
\label{D1} \Delta_{out} &=& -1 + \left(\frac{A}{3a^{3\gamma}} +
\frac{V_F}{3}\right)B^2 \,,\\
\nonumber \\
\Delta_{in} &=& -1 + \frac{V_T}{3}\,B^2 \,.\label{D2}
\end{eqnarray}

Now we proceed to write the equations for the evolution of the
bubble in outside coordinates. In order to do that we rewrite
Eq.~(\ref{D0}), by using Eqs.~(\ref{D1}) and (\ref{D2}), obtaining
\begin{equation}\label{radio-burbuja}
\dot{B}^2 = B^2\,C^2 - 1,
\end{equation}
where
\begin{equation}\label{radio-burbuja2}
C^2 = \frac{V_T}{3} + \left(\frac{\sigma}{4} +
\frac{1}{\sigma}\left[\frac{V_F - V_T}{3} +
\frac{A}{3a^{3\gamma}}\right]\right)^2.
\end{equation}

In the outside coordinates we parameterize $x(t)$ as the curve for
the bubble evolution (the bubble radius in these coordinates). Since
$x$ and $t$ are dependent variables on the shell, this is
legitimate. We write $B = ax$, then by using $\dot{B} =
a_{,t}\,x\,\dot{t} + a\,\dot{x}$ and
\begin{equation}\label{dt}
\frac{dt}{d\tau} = \frac{1}{\sqrt{1 - \frac{a^2}{\left(1 -
x^2/R^2\right)}\left(\frac{dx}{dt}\right)^2}},
\end{equation}
obtained from Eq.~(\ref{cond1}), we can express
Eq.~(\ref{radio-burbuja}) as follow
\begin{equation}\label{ecuacion-x}
\frac{dx}{dt} = \pm \sqrt{\frac{\left(R^2-
x^2\right)\left(a_0^2\,C^2\,x^2
-1\right)}{x^2\,a_0^2\left(a_0^2\,C^2\,R^2 - 1\right)}}\,.
\end{equation}

The evolution of $\sigma$ is determinate by Eq.~(\ref{cons2}) which
could be converted to outside coordinates by using Eq.~(\ref{dt})
obtaining
\begin{equation}\label{ecuacion-sigma}
\frac{d\sigma}{dt} =
-2\left(\frac{\bar{\gamma}\,\sigma}{x}\right)\frac{dx}{dt} +
\frac{a_0\,\gamma\,\rho_0}{\sqrt{-\left(\frac{dx}{dt}\right)^2a_0^2
+ 1 -\frac{x^2}{R^2}}}\,\frac{dx}{dt}\,.
\end{equation}

The positive energy condition $\sigma > 0$ together with
Eq.~(\ref{D0}) impose the following restriction to $\sigma$
\begin{equation}
 0 < \sigma \leq 2 \sqrt{\frac{V_F-V_T}{3} +
\frac{\rho_0}{3}}\;.
\end{equation}

Also, from the definition of $x$ and Eq.(\ref{ecuacion-x}) we obtain
the following restriction for $x$
\begin{equation}
\frac{1}{a_0C} \leq x \leq R \;.
\end{equation}

We solved the Eqs.~(\ref{ecuacion-x}, \ref{ecuacion-sigma})
numerically by consider different kind and combinations of the
matter content of the background and the bubble wall.
From these solutions we found that once the bubble has materialized
in the background of an ES universe, it grows filling completely the
background space.

In order to find the numerical solutions we  chose the following
values for the free parameters of the model, in units where $8\pi G
= 1$:
\begin{eqnarray}
a_0 &=& 1 \,,\\
V_T &=& 0.1 V_F \,,\\
\sigma_{init} &=& 10^{-6}\,.
\end{eqnarray}

The other parameters are fixed by the conditions discussed in
Sec.\ref{Sec-static}.


\begin{figure}
\centering
\includegraphics[width=7cm]{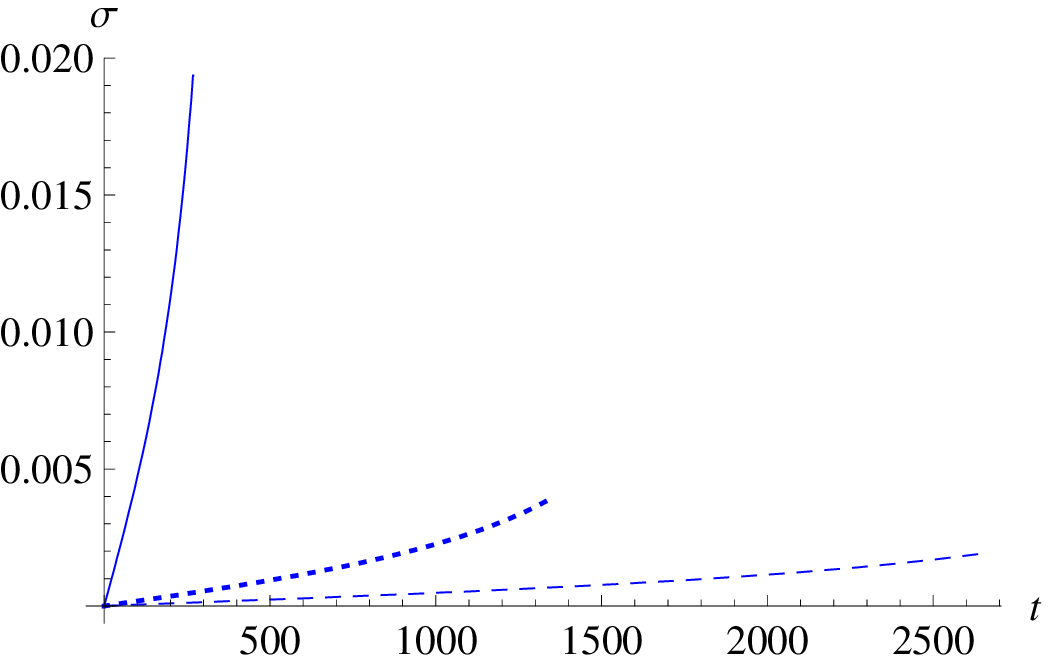}
\includegraphics[width=7cm]{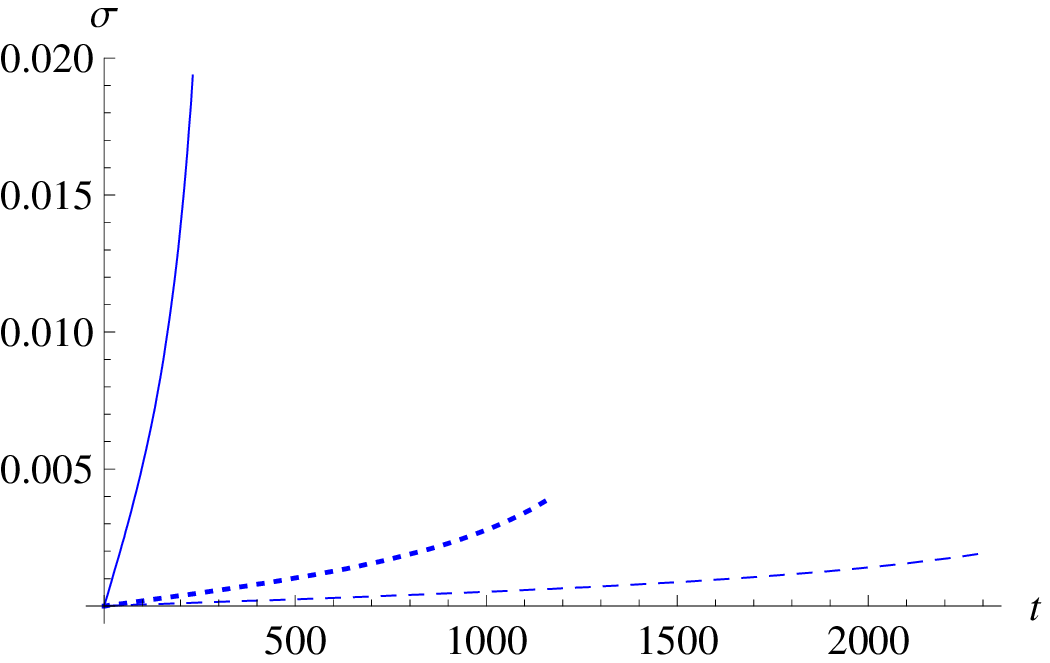}
\includegraphics[width=7cm]{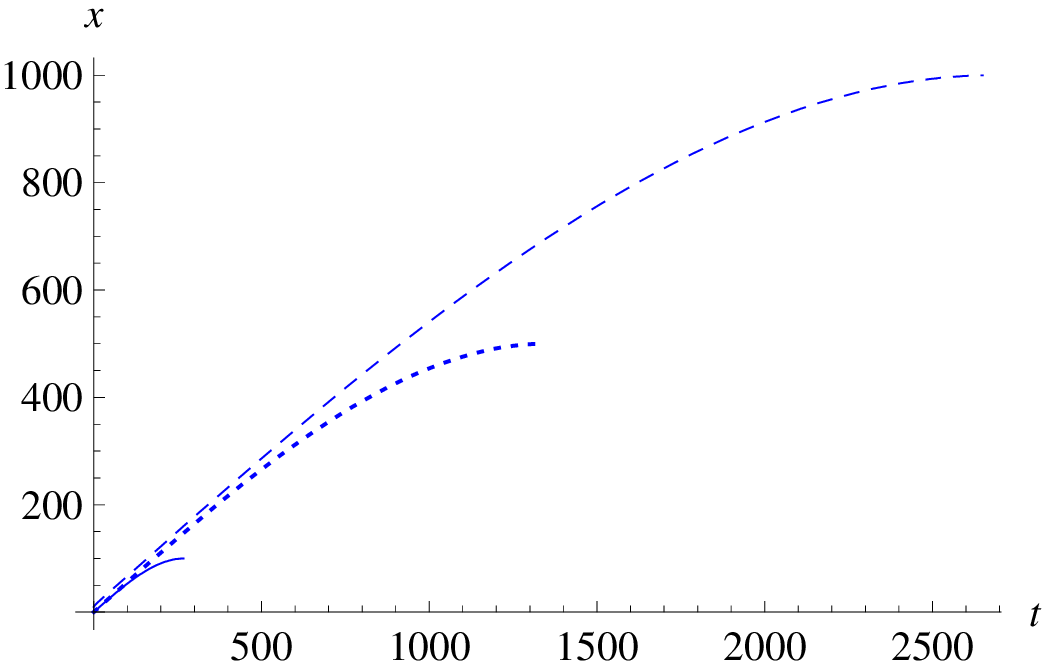}
\includegraphics[width=7cm]{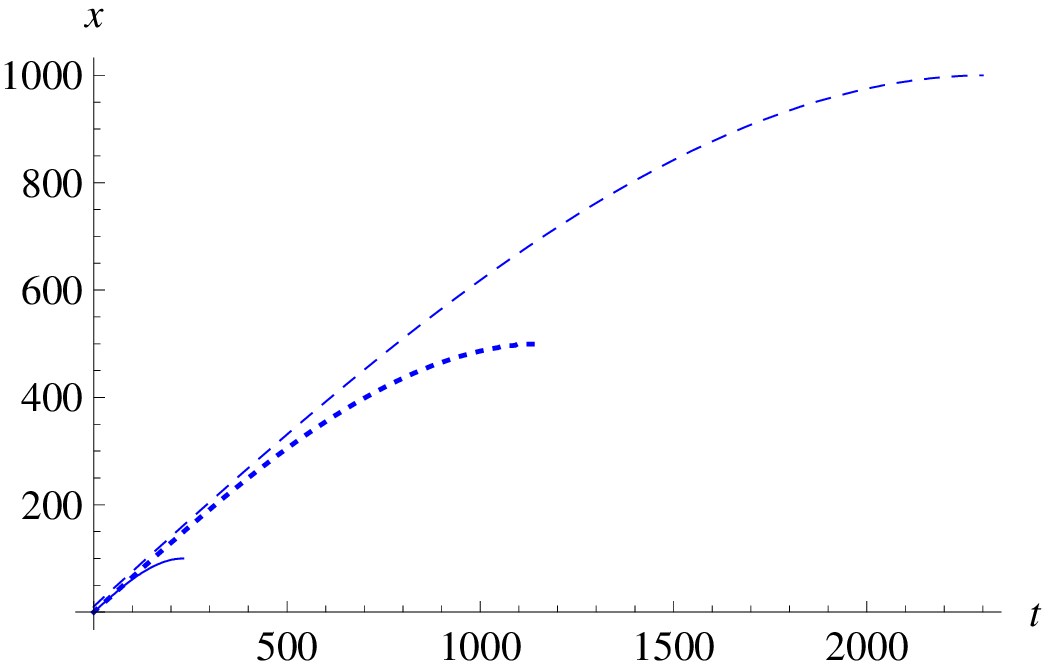}
\caption{Time evolution of the bubble in the outside coordinates
$x(t)$, and time evolution of the surface energy density
$\sigma(t)$. The left panel is for a static universe dominated by
dust and the bubble wall containing dust. The right panel is the
same situation but with radiations instead of dust. In all these
graphics we have considered dashed line for $R=1000$, dotted line
for $R=500$ and continuous line for $R=100$.\label{Soln1}}
\end{figure}


\begin{figure}
\centering
\includegraphics[width=7cm]{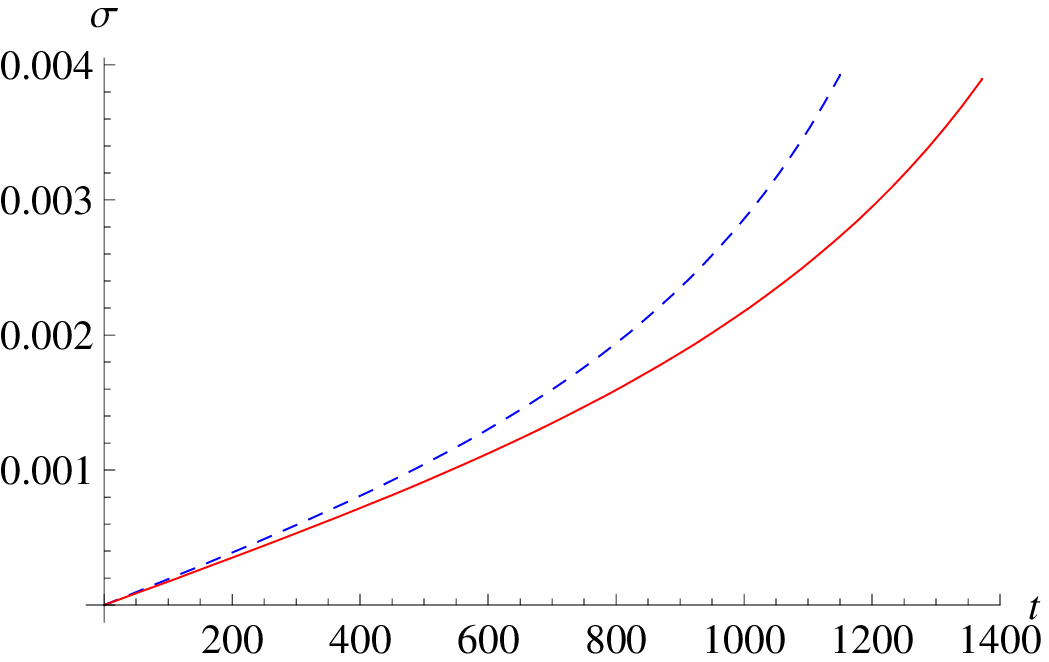}
\includegraphics[width=7cm]{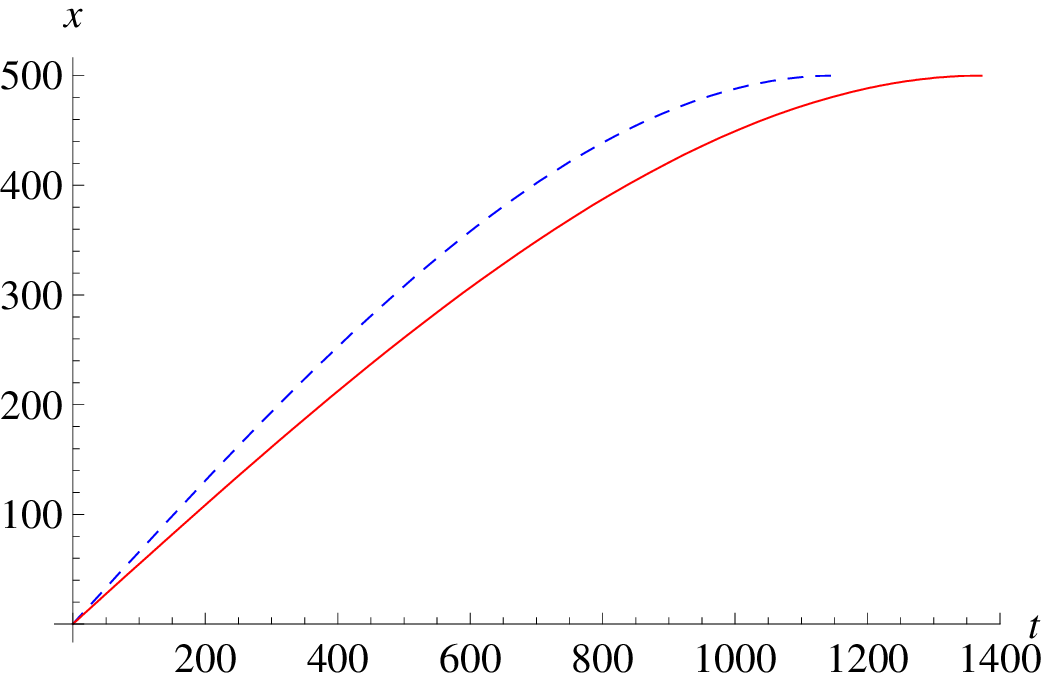}
\caption{Time evolution of the bubble in the outside coordinates
$x(t)$, and time evolution of the surface energy density
$\sigma(t)$, for a background with $R=500$. Dashed line corresponds
to a static universe dominated by dust and bubble wall containing
radiation. Continuous line corresponds to a static universe
dominated by radiation and a bubble wall containing
dust.\label{Soln2}}
\end{figure}


Some of the numerical  solutions are shown in
Figs.~(\ref{Soln1},\ref{Soln2}) where the evolution of the bubble,
as seen by the outside observer, is illustrated. In these numerical
solutions we have considered three different curvature radius
($R=1000$, $R=500$, $R=100$) and various matter contents
combinations for the background and the bubble wall.
From these examples we can note that the bubble of the new face
grows to fill the background space, where the shell coordinate
asymptotically tends to the curvature radius $R$.

\section{Conclusions}\label{Sec-conclusions}

In this paper we explore an alternative scheme for an Emergent
Universe scenario, where the universe is initially  in a truly
static state. This state is supported by a scalar field which is
located in a false vacuum. The universe begins to evolve when, by
quantum tunneling, the scalar field decays into a state of true
vacuum.

In particular, in this work we study the process of tunneling of a
scalar field from the false vacuum to the true vacuum and the
consequent creation and evolution of a bubble of true vacuum in the
background of Einstein static universe.
The motivation in doing this is because we are interested in the
study of new ways of leaving the static period and begin the
inflationary regime in the context of Emergent Universe models.

In the first part of the paper, we study a Einstein static universe
dominated by two fluids, one is a standard perfect fluid and the
other is a scalar field located in a false vacuum. The requisites
for obtain a static universe under these conditions are discussed.
As was shown by Eddington \cite{Eddington}, this static solution is
unstable to homogeneous perturbations, furthermore it is always
neutrally stable against small inhomogeneous vector and tensor
perturbations and neutrally stable against adiabatic scalar density
inhomogeneities with high enough sound speed \cite{Gibbons:1987jt,
Gibbons:1988bm, Harrison:1967zz, Barrow:2003ni}. This situation has
implication for the EU scenario, see discussion bellow.

In the second part of the paper, we study the tunneling process of
the scalar field from the false vacuum to the true vacuum and the
consequent creation of a bubble of true vacuum in the background of
Einstein static universe. Following the formalism  presented in
\cite{KeskiVakkuri:1996gn} we found the semiclassical tunneling rate
for the nucleation  of the bubble in this curved space.
We conclude that the probability for the bubble nucleation is
enhanced by the effect of the curvature of the closed static
universe background.

In the third part of the paper, we study the evolution of the bubble
after its materialization. By following the formalism developed by
Israel \cite{Israel:1966rt} we found that once the bubble has
materialized in the background of an ES universe, it grows filling
completely the background space.
In particular, we use the approach of \cite{Fischler:2007sz} to find
the equations which govern the evolution of the bubble in the
background of the ES universe. These equations are solved
numerically, some of these solutions, concerning several type of
matter combinations for the background and the bubble wall, are
shown in Figs.~(\ref{Soln1},\ref{Soln2}).

In resume we have found that this new mechanism for an Emergent
Universe is plausible and could be an interesting alternative  to
the realization of the Emergent Universe scenario.

We have postpone for future work the  study of this mechanism
applied to Emergent Universe based on alternative theories to
General Relativity, like Jordan-Brans-Dicke \cite{Jbd}, which
present stable past eternal static regime
\cite{delCampo:2007mp,delCampo:2009kp}.
It is interesting explore this possibility because emergent universe
models based on GR suffer from instabilities, associated with the
instability of the Einstein static universe.
This instability is possible to cure by going away from GR, for
example, by consider a Jordan Brans Dicke theory at the classical
level, where it have been found that contrary to general relativity,
a static universe could be stable, see \cite{delCampo:2007mp,
delCampo:2009kp}.
Another possibility is considering non-perturbative quantum
corrections of the Einstein field equations, either coming from a
semiclassical state in the framework of loop quantum gravity
\cite{Mulryne:2005ef,Nunes:2005ra} or braneworld cosmology with a
timelike extra dimension \cite{Lidsey:2006md, Banerjee:2007qi}. In
addition to this, consideration of the Starobinsky model, exotic
matter \cite{Mukherjee:2005zt, Mukherjee:2006ds} or the so-called
two measures field theories \cite{delCampo:2010kf,delCampo:2011mq,
Guendelman:2011fq,Guendelman:2011fr} also can provide a stable
initial state for the emergent universe scenario.

On the other hand, in the context of GR the instability of the ES
could be overcome by consider a static universe filled with a
non-interacting mixture of isotropic radiation and a ghost scalar
field \cite{Barrow:2009sj} or by consider a negative cosmological
constant with a universe dominated by a exotic fluid satisfies
$P=(\gamma -1)\rho$ with $0 < \gamma < 2/3$, see
\cite{Graham:2011nb}. In this case it is important that the exotic
matter source should not be a perfect fluid. It could be, for
example, an assembly of randomly oriented domain walls
\cite{Bucher:1998mh}.

We are interested in apply the scheme of Emergent Universe by
Tunneling developed here to models which present stable past eternal
static regimes, in the near future.

\section{acknowledgments}

P. L. is supported by FONDECYT grant N$^{0}$ 11090410.
PL wishes to thank the warm hospitality extended to him during his
visits to Institute of Cosmology and Gravitation, University of
Portsmouth were part of this work was done.
We are grateful to R. Maartens for collaboration at the earlier
stage of this work and to A. Cid  for reading and comments about the
manuscript.

\end{document}